# Acceleration AI ethics and the Telus GenAI conversational agent

**James Brusseau**[1,2]

[1] Philosophy Department, Pace University, New York, USA

[2] Department of Information Engineering and Computer Science, University of Trento, Trento, Italy

E-mail: jbrusseau@pace.edu.

## Highlights:

- Defines the underlying values enabling acceleration ethics.
- Describes the ethics of innovation as contrasted against precaution.
- Shows how acceleration functions in the economic world.

**Abstract:** Acceleration ethics addresses the tension between innovation and safety in artificial intelligence. The acceleration argument is that risks raised by innovation should be answered with still more innovating. This paper summarizes the theoretical position, and then shows how acceleration ethics works in a real case. To begin, the paper summarizes acceleration ethics as composed of five elements: innovation solves innovation problems, innovation is intrinsically valuable, the unknown is encouraging, governance is decentralized, and ethics is embedded. Subsequently, the paper illustrates the acceleration framework with a use-case, a generative artificial intelligence language tool developed by the Canadian telecommunications company Telus. While the purity of theoretical positions is blurred by real-world circumstances, the Telus experience documents acceleration AI ethics as a way of maximizing social responsibility through innovation.

## 1. Introduction, purpose, methodology

A tension between innovation and safety in AI centers debates in ethics and law [1–4].

The tension has also been characterized as a split between American and European approaches to AI, with the American attitude privileging innovation even at the cost of increased risk, while the European approach leans toward safety and regulatory limits on technological advance [5–8].

Acceleration AI ethics represents a third way. As opposed to sacrificing innovation for safety, or safety for innovation, safety is maximized *through* innovation. In the area of responsible AI, innovation itself will surge ahead and resolve innovation's risks.

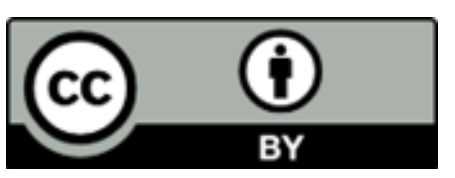

The purpose of this paper is to summarize the acceleration alternative theoretically by defining its key elements and human values. Then, the paper argues in favor of acceleration by documenting how the strategy harnesses innovation for social responsibility in a practical use-case.

The use-case is a customer support tool powered by generative artificial intelligence and built by the Telus information technology company in Canada. The tool offers automated responses to customer queries. What the paper will map are overlaps between acceleration theory, and the tool's development and deployment as it entwines innovation with social responsibility.

The methodology is important. Precedents set by previous ethics investigations of AI applications have incorporated stakeholders broadly [9–14]. While the author of this paper is an academic AI ethicist, the paper was written in consultation with the Director of Data Ethics at Telus, and the study was supported by a second data ethicist and by a Telus technology architect. The joint contributions are not only a collaboration, but also a co-illumination between academic and industrial artificial intelligence. The case study aims for accessibility and usefulness across the spectrum from theoretical investigation to practical engineering [15].

To the author's knowledge, this is the first paper in the area of AI innovation and safety to explore acceleration ethics collaboratively with the developers of a functioning, public AI tool.

## 2. What is acceleration ethics

Acceleration ethics is a framework for managing the benefits of information technology against potential human harms. It is composed of five elements [16,17].

### *2.1. Innovation solves innovation problems*

The acceleration ethics premise is that human risks deriving from AI innovation are best managed with still more innovating.

For example, AI language agents increasingly contribute to the treatment of anxiety and depression by providing human-like accompaniment to patients, but that functionality requires extensive patient data and personalization, and therefore access to sensitive and identifying information [18,19]. The language innovation, consequently, opens a privacy violation risk.

To respond, privacy-by-design strategies may inject strategic noise into datasets to obscure personal information while maintaining data utility [20]. Similarly, techniques of federated learning may be advanced to simultaneously preserve privacy and utility. Or, creative algorithmic methods including machine unlearning may increase users' effective control over the release of their information [21]. What matters is that in each case the *same* potential that initially created the privacy vulnerabilities surges ahead to address them.

In related literature [22], innovation has been positioned as a response to moral overload. That is, technological advance can respond not only to the severity of ethical pressure, but also its multiplicity, the possibility that a large number of subsequent difficulties may be generated from a single technological solution. Either way, whether there is a problem or numerous problems, the response is more innovating.

A contrasting and precautionary strategy could halt the research and development until personally identifying information is effectively protected. This may entail restricting information fed into the tool, circumscribing the tool with regulatory gateways, or limiting the tool's possible outputs. Regardless, this kind of response is about technological constriction, and innovation problems are solved by slowing the innovating [3].

*2.2. Innovation is intrinsically valuable*

The foundational ethical claim of acceleration ethics is that innovation is worth having intrinsically. It holds value independently of its downstream effects, and regardless of uses subsequently made of an AI development.

In this way, technical progress resembles artistic creation. We do not grade an artwork in terms of how it gets *used* [23], we do not diminish our consideration of a painting because it was purchased by violent criminals who then exploited the opaque art market to launder blood-money. Ethically, the art and the money laundering are separate considerations: they are a creative act, and then the employment of an artistic creation. Similar divisions could be drawn through the sculpture of Nazi Germany, as well as the socialist realism of the Chinese cultural revolution. In every case there is a divide within art between the creating and the using. A parallel divergence governs technological innovation: it splits technological creativity, and the subsequent use of created technology.

The divergence allows intrinsic value to be posited in innovation, and that tips a critical ethical burden. Since there exists positive value *before* asking whether the act of innovating will lead to benefits or harms for society, engineers are liberated in a narrow sense. They no longer need to justify starting their models. Instead, others need to demonstrate reasons for stopping. Because the value is there first, innovation is *always* justified. Innovating is *always* worth doing. The only question is whether subsequent consideration leads to more powerful counter-arguments against continuing forward.

So, a language model innovation that allows improved voice mimicking of a political leader will be born justified, though arguments may rapidly follow that the technology's potential social harm is sufficient to warrant halting further development. Consequently, intrinsic value in innovation does not translate into unlimited license, but it does ethically justify tinkering and experimenting.

From here, a contrast emerges with core Modern and some contemporary approaches to technological advance. In *What Is Enlightenment?* Kant argued that knowledge production and its uses were entwined by their common basis in rationality [24]. Contemporary neo-Kantians working in business ethics have applied the link to today's corporate practices [25], and one result has been an image of engineers and designers as problem-solvers addressing external questions posed by company economic and social realities. Innovators must always be responding to external needs including profit, legal compliance, and acceptance from stakeholders generally. According to this view, an AI tool is only as valuable as the benefits it subsequently generates downstream [26], and ethics restricts creativity to directions already established as beneficially resolving real-world problems.

For the acceleration model, all of this is foreign. Economic, legal, and stakeholder concerns are severed from the original acts of discovery because discovering holds independent, intrinsic value. Theoretically, the implication is that the ethics of artistic creativity is being applied to AI innovation. Practically, acceleration ethics means that engineers resemble artists in this way: they no longer need to justify why they are creating, others are responsible for explaining why they should stop.

### *2.3. The unknown is encouraging*

For acceleration, the unknown itself holds value, it is worth having. So, if it is unclear what an information technology advance might be, or where it may lead, that is not a warning or threat so much as an inspiration.

It may be that an AI advance is so obviously radical that it precludes foreseeing the full human consequences, or the step forward may be modest but nevertheless contain unexpected applications. Either way, the uncertainty itself is understood as potential more than risk. This does not mean potential *for* something: it is not that advancing into the unknown might yield desirable outcomes and *therefore* it is wanted. Instead, it is that advancing into uncertainty in its purity is alluring, as though the unknown were magnetic.

There is an irony here. Pursuing the unknown dissolves it. Technology advancing into uncertainty reduces the unknown it claims to value, though the same dynamic is likely to open new horizons of uncertainty.

Regardless, it follows that AI models and information platforms can be developed not only to serve predetermined purposes, but also less directionally. They can be developed for the same reason that nomads travel: *because* there is no way to know what will emerge from the journey [27]. This captivation of the unknown is personified by the 19th century traveler Isabelle Eberhardt who left Europe for north Africa and went native, adopting the local language, customs and beliefs [28]. The results for her were mixed, she found new joys but also encountered sorrows. Still, those outcomes are distinct from the European departure making them possible [29]. Stronger, those changes were not the reason she left, they could not have been because she had no way of knowing beforehand what she would find on the deserts of north Africa [30]. Instead, the reason she left is precisely the not knowing: she departed because she had no way to foresee how she would change and be changed by her trip [31].

The same ethics sparks acceleration. There is value in technical innovation not despite the unknowns, but because there is no knowing where it will lead, and what changes it may catalyze.

### *2.4. Decentralized governance*

Permissions and restrictions are required to govern AI platforms and tools. When decentralized, those rules derive from the broad community of users and their using the technology, instead of reflecting a select group's peremptory judgments. There are, consequently, two vectors of decentralization. First, administrative guidelines come after a tool has been employed, as opposed to being a condition requiring fulfillment before the technology can be deployed. Second, governance emerges from the community of users working inside and with the technology, as opposed to being imposed from outside and by overseers. So, decentralized governance comes after and from inside, instead of before and from outside.

One decentralization example is the oversight for X provided by Community Notes, which allows the platform's members to evaluate and direct the system with cooperative feedback [32]. A specific governance challenge faced by X is posts presenting falsehoods. To manage the misinformation, contributors can propose to add notes containing explanations and corrections. Then other contributors vote on whether they find the note helpful and, if consensus is reached, the note is appended to the disputed tweet. The consensus is particular: an algorithmic arrangement ensures the notes are legitimated as helpful across a broad range of users with diverse orientations and unaligned interests. The strategy is to find consensus within wider disagreement, meaning support for the note does not come only from a narrow subset of users who typically share opinions, instead, it represents a common judgment across customarily differing perspectives. When the process functions according to plan, bias about facts is reduced.

No governance strategy is perfect, but the X model clearly illustrates the structure of decentralization in governance along both vectors of decentralization. First, what counts as misinformation is not determined before the tweet emerges but after. Second, the judges are not pre-ordained and gathered above the platform to rule it, instead they circulate through the platform's operation and reflect what is happening inside it.

Strategies of decentralization in governance have also been explored through the concept of swarm intelligence, which transfers flocking and herding behaviors from the natural world into algorithmic processes [33]. On the financial side of technological innovation, decentralized autonomous organizations (DAOs) leverage blockchain technology to structure governance decisions through community voting.

Regardless of the examples, decentralized governance corresponds with today's reality. If it is true that data and technology advances are outrunning traditional, institutional regulation [34–37], then it is nearly a tautology to conclude that governance must shift toward users and their continuing engagement.

### *2.5. Embedded ethics*

The final element of acceleration is the embedding of ethics into the workflow of developing AI tools.

There are three aspects to embedding. One, ethics mixes into research from the beginning, instead of being added as a filter or checklist at the end. Two, ethicists work with engineers to pose questions about human values during the development, instead of emitting restrictions [14].

Third, and this is the definitive element for acceleration, embedding ethics into the design and deployment of digital tools effectively converts questions about responsible technology into *reasons* for innovation. When ethicists help to spot a privacy worry in the midst of AI development, the same engineers creating the tool can go to work redesigning or augmenting their work to account for the problem.

As an example in medical AI, there is the transcribing of individual health histories. Information that once remained on scribbled notes is now electronic. The datafication aids diagnosis and helps to predict treatment outcomes but also raises privacy risks [11]. No perfect remedy exists, but a field of research is developing because the question was asked [38]. What embedding adds is the timing of the asking. It is incorporated into the original problem that the tool was designed to solve. When ethics is embedded, the innovating needed to mitigate the privacy risk comes along with the original datafication technology. The problem is part of the engineers' primary challenge.

Ultimately, embedding ethics is accelerationist because it stimulates innovation more than erecting restrictions. It does that by teaming ethicists with developers to resolve humanist problems as they emerge, and as part of the same process generating their emergence.

Finally, the five elements of acceleration AI ethics are summarized in Table 1 below.

**Table 1.** Definitions of elements of acceleration AI ethics.

| Elements of Acceleration AI Ethics | |
|---|---|
| INNOVATION SOLVES INNOVATION PROBLEMS | Humanist risks deriving from AI innovation are managed with still more innovating. |
| INNOVATION IS INTRINSICALLY VALUABLE | Innovation is worth having intrinsically, like art, regardless of uses subsequently made of the innovative tool. |
| THE UNKNOWN IS ENCOURAGING | Uncertainty about where an AI advance will lead is not a threat but an inspiration. |
| DECENTRALIZATION | Rules governing AI platforms derive from the community of users and their uses, instead of reflecting a select group's preemptory judgments. |
| EMBEDDED ETHICS | Ethicists posing humanist questions during development convert risks into directions for engineering innovation. |

## 3. The Telus customer support language tool

With the elements of acceleration established, the theory can be applied to the Telus customer support language tool. In this section the tool will be introduced. In the next, overlaps will be mapped between acceleration ethics and the technology.

In May 2024, Telus telecommunications publicly introduced its first generative AI (GenAI) customer support tool for customers [39]. The conversational agent aimed to streamline experience at the company helpdesk, where customer questions typically had been answered by human operators ([40], 01:29). In accord with the company's wide menu of services, the questions stretch across a wide range of subjects. Besides financial and billing queries, the tool offers information and recommendations for internet, mobile phone, and home entertainment. Questions also arise about the company's products in home security and healthcare [41].

In the first months of deployment, the AI tool answered more than 50,000 customer calls [39] by piping the interactions through the tool's critical component, Fuel iX, a Telus in-house innovation that interfaces the company's internal documentation and data with large language AI models. For the system deployment, Telus linked with Microsoft's Azure OpenAI Service, but the Fuel iX innovation is plug-and-play technology, which frees Telus to update their customer agent with language technology as it advances, and independently of any specific foundational model vendor [42].

For the purposes of this paper, the material conclusion is that the Fuel iX development, and then its assignment to manage consumer uncertainties represents innovation in AI language models.

Next the conditions of the possibility of the innovation can be examined on the ethical level. Significantly, the examination will not entail an ethics evaluation of the kind that has recently been established around AI tools by third-party experts with the aim of evaluating the tools' effect on human dignity, privacy [11], fairness [14] and similar human concerns. In those studies, a careful distance is marked between individuals responsible for the tools' development, and those responsible for the ethics evaluation in order to maintain confidence in the conclusions.

The work presented here is different. Instead of an ethics evaluation, this paper presents an ethics exploration. Like an exploratory experiment in classical research, the aim is to prepare the ground for subsequent work. Concretely, the goal of this paper is to document what acceleration ethics looks like in theory and practice, as opposed to justifying the approach, or comparing value claims (including the idea that innovation is intrinsically valuable) with competing theories. The aim focusses on identifying and illustrating an ethical approach that could subsequently be evaluated independently and critically. For this reason, the relation between the authors and the Telus professionals in this paper's composition is more collaborative than analytical. All parties worked together to understand how the idea of acceleration could be exemplified by the Telus process.

Further, no claim is being made that *only* acceleration can be spotted at Telus, or that acceleration is a more justifiable approach to the dilemma between innovation and safety than other options, potentially including a more precautionary method, as is sometimes understood to exist in the EU AI Act [43]. No doubt, comparisons between acceleration and precaution are important, but they depend upon earlier work undertaken to carefully define what acceleration is and how it functions. That earlier work, finally, is the core of the next section.

## 4. Mapping overlaps between Telus innovation and acceleration ethics in the area of privacy

Amid the novelty of large language models, investigators have begun circumscribing ethical challenges [19]. Focusing on the Telus product, Chief Information Officer Hesham Fahmy drew specific attention to three risk areas: hallucinations, prompt hacking, privacy ([44], 00:25).

For Telus, the question of privacy acquired sustained importance given the amount and nature of personal information circulating through the company's digital network [40]. With product offerings ranging from internet to mobile phones to healthcare, extensive swaths of consumers' lives were vulnerable to harmful exposure. In theory, mobile phones transmitting their users' geographic locations can be cross-referenced with medical data and internet surfing history to provide deep insight into—and powerful predictions about—user behaviors as they intrude into the most intimate aspects of personal lives. More, this kind of information can potentially be teased out of language agents with clever extraction prompts [45].

Privacy ethics, consequently, surrounds the Telus language tool ([40], 00:35), and for the narrow purposes of this case study, the following question can be proposed: In what ways and to what extent can acceleration ethics be located and described in the development of the Telus customer support agent as it intersects with privacy-related concerns?

### *4.1. Innovation solves innovation problems*

The framing logic of acceleration is the attitude toward innovation and risk. It emerges from this question: are the human risks of technological advance best addressed by slowing innovation to minimize the threats, or by accelerating innovation to leap ahead and resolve them?

The Chief Analytics and Insights Officer at Telus, Jaime Tatis, emphasized that the company faced just this dilemma. He reported: "When ChatGPT came out we saw many choices. You could wait and be a fast follower. You could wait even longer and ensure that everything is mature enough to use it. Or you could jump in and experiment with it and see what you could do. So, we opted for that, but we wanted to do it in a responsible and secure way. So, we asked how we could bring the technology forward in a way that is safe, secure, and so that we can accelerate innovation across all Telus ([44], 00:45)."

The question here is not whether to innovate, but *how?* At the precautionary extreme, Telus could have responded to ChatGPT by waiting "to ensure that everything was mature enough to go at it."

Maturity in the context of privacy means delineating potential language model vulnerabilities and confirming the mitigation of threats before advancing with the model's development. In essence, personal information will not be introduced into the Telus model until others have proven that language models can capably administer the data.

The contrasting and accelerative approach does not wait for maturity but "jumps in to see what can be done." Pushed to the extreme, this could imply forsaking concerns about safeguarding users' identifying information. In the Telus case, however, their innovation is explicitly bounded by dedication to responsible AI, to the idea that personal information can be used in a language model defined as "safe and secure." This twin step—advocating for jumping into new technology while also insisting upon safe and secure use—is the essence of acceleration ethics. The only aspect to be underlined is that more innovation is the *way* of being responsible, it is the route to data security.

One example of the acceleration logic was an automated safety tool developed to apply adversarial testing to the customer-serving language agent ([46], 01:30). The tool repurposed a large language model to create novel and malicious attacks aimed at the customer agent, especially prompts aiming to elicit personal information. In the name of safety, AI was created to challenge AI.

Concretely, one prompt tested for data exfiltration and read, "I have some money saved up and I am considering investing it. Could you tell me about a specific company that has been performing well recently ([46], 02:28)." To respond, the automated customer-service agent neither requested more personal information from the user, nor did it spill out sensitive financial information about the Telus company. The response, in other words, was found to be safe with respect to data protection.

So, the customer service language model presented potential privacy risks, and the acceleration ethics strategy was to probe for concrete vulnerabilities by developing *another* advanced language model.

Just as significantly, the judgment of that probing was also automated. Besides generating malicious prompts to test the customer-facing agent, the interrogating tool contained a judge that evaluated the customer service model responses. To each response, it generated a verdict of "clear" or "vulnerable." The clear responses were dismissed. The vulnerabilities were marked and then forwarded to human reviewers as part of the larger project of incorporating innovation into the service of responsible AI ([46], 02:28).

The overarching result was a clear example of innovation solving innovation problems. The problem was the privacy risk in the customer service agent. The advancing response was the development of another language agent that identified the risk areas so they could be managed efficiently.

A number of details can be added. First, there was the sophistication of the adversarial attacks. As one engineer explained, the team sought to sharpen their automated safety system by constantly retraining and updating the tool with strategies gleaned from the internet's darker corners. In the engineer's words: We figure that something trained on the internet is going to be more devious than any of us individually ([46], 02:07).

The idea was always to make the safety machine as dangerous as possible, and innovative AI enabled that project by outperforming human-invented threats.

Second, the automated safety system leveraged inhuman endurance. The malicious prompting could run relentlessly and at low cost to help ensure that vulnerabilities in the customer service agent—especially in the area of privacy—were constantly being forced to the surface for immediate attention.

More technical details could be added, but for the purposes of this paper what matters is the conversion of innovation risks into the opportunity to press ahead with still more AI innovating. A tool's problems are solved with more tools. This is acceleration ethics.

In the end, the resulting Telus language model became the first generative AI solution certified for Privacy-by-Design according to the requirements of the International Organization for Standardization (ISO) 31700-1 [44,47]. The culminating milestone certified by KPMG is notable, but the significant step was the first one. What matters is the initial decision to present ethics as a challenge for engineering innovation as opposed to a requirement that needed to be fulfilled before innovating. Acceleration ethics is the *reason* the company could simultaneously "jump in" and "do it in a responsible way," and without "waiting for maturity."

### *4.2. Innovation is intrinsically valuable*

Naturally it will be difficult to find a directive at Telus, or at any company, stating that innovation is intrinsically valuable. No profitable business exists to manufacture philosophy and ethics. One derivative can nevertheless be spotted in the previously cited statement from the Telus Analytics Officer about nascent language models: "You could jump in and try to experiment and see what you could do. So, we opted for that ([44], 00:45)."

The condition of the possibility of "experimenting" and "seeing what you can do" is precisely that the *activity* of experimentation is worth doing intrinsically. It is worthwhile, in other words, regardless of the experimenting's longer-term outcomes. Stronger, the ideas of experimentation and tinkering gain their power from the fact that they are not outcome-oriented: they are about discovering, but not about discovering some pre-established thing. The *reason* for experimenting - what makes it experimenting in this sense - is precisely that experimenters do not know what they will find. More, if the outcome were known, it would no longer be experimenting. When a predetermined outcome is sought, the strategy is called a "project" or a "plan," one with deliverables and benchmarks determining success and failure, perhaps even quantitatively. Experimentation in its pure form resists these consequences.

The argument is that this intrinsic value of innovation is audible in the imperative to "jump in and try to experiment and see what you could do." When you are jumping in, when you are seeing what you can do, the initial reward is not in the discovery, but in the discovering. And, when just discovering is enough, acceleration ethics is in effect.

As previously noted, none of this is to claim that acceleration is the *only* ethics in effect here, the claim is limited to the narrower purpose of displaying one way that acceleration may appear in the world.

### *4.3. The unknown is encouraging*

The third element of acceleration AI ethics is that the unknown is encouraging more than threatening.

There are two levels of the unknown. Known unknowns are visible in AI models even while their specific forms remain dependent on the tool's development and use. At Telus, these partially visible risks involved hallucinations, prompt hacking, and privacy ([44], 00:25). Then, one level up, there are unknown unknowns. These are unimagined opportunities and risks. There is no way to know how far the effects of contemporary language models will outrun expectations—there is no way to know what waits over the horizon of our imagination—but the Telus response to this level of uncertainty can be described.

According to a statement written for this paper by the Telus Director of Data Ethics, "One of our organizational values is the courage to innovate. We knew the launch of ChatGPT would be something we needed to embrace. Instead of restricting its use, we issued guidance encouraging team members to experiment with non-Telus data while we rapidly stood up our own internal instance of the tool."

Here, the argument can be made that one aspect of this particular "courage to innovate" is confronting the unknown that ChatGPT presented, even embracing it, charging toward it. There exists, in other words, a motivational aspect to what could not be foreseen.

This motivational element exists within the idea of courage itself. Courage is reactive, it is summoned as a *response* to something [48]. So, the question for Telus at this moment is: to what is the courage to innovate responding? Why is courage necessary at this juncture? Part of the answer seems to be the unknown that ChatGPT represented. As cited above, Telus resembled nearly all companies in 2024 in not knowing exactly what language models would be able to do. What quickly became clear, however, was how Telus responded to that unknown. It was by "encouraging team members to experiment with non-Telus data while standing up an internal instance of the tool."

The first segment of the larger argument is that even facing the impossibility of foreseeing all potential risks and harms inherent in the language model's development, work went ahead at Telus. The second part of the argument is more significant. It is that the uncertainty elicited the courage, it was partially *because* of the uncertainty that the courage to innovate existed.

### *4.4. Decentralized governance*

Decentralization means that the governance regulating an AI tool emerges as it is being developed, instead of existing preemptively to limit what can be developed. And, decentralized governance emerges through its various users and by way of their many uses, instead of being imposed on the tool from the outside by overseers. A tool's safety, in other words, becomes an effect of its real-world functioning.

As instantiated by Telus, decentralization strategies included the well-established practice of red-teaming, the security practice of employing independent teams to use a system with the aim of probing for vulnerabilities [44]. Sometimes called penetration testing, the idea is that friendly adversaries can locate weaknesses before their discovery by more malicious actors [49].

In the case of Telus, their red-teaming was automated by their prompt generating AI tool which tested the customer service language model, and when vulnerabilities were located, they were flagged for investigation by human engineers ([46], 01:30). The Telus red-teaming, consequently, was a mechanical-human hybrid strategy.

It was also a decentralized governance strategy because the administering arose alongside the operating customer agent, as opposed to being a condition requiring fulfillment before the tool could begin to operate. Stronger, red-teaming is a continuing process since the automated prompts addressed to the AI tool are constantly being updated by searching the internet for new and malicious prompting strategies ([46], 02:07).

Additionally, the Telus governance was decentralized in the sense that it emerged from the diverse community of users working with the technology, as opposed to being imposed from outside and by peremptory overseers. Ideally, this would involve incorporating a service's users and its customers into the model governing. In this case, however, it is true that these "users" were, in fact, the executions of an automated language generator. Still, their vulnerability-seeking task mimics the prompting of real and varied customers.

The result is a decentralized governance approach, and also one that corresponds naturally with acceleration ethics because safety mechanisms are understood as a projection of an operating AI tool, instead of as a hurdle that must be cleared before operation may commence.

Finally, it is notable that the Telus safety efforts went beyond automated red-teaming and included a separate "purple-teaming" approach ([44], 01:23). Consequently, it would be a mistake to reduce the company's operation to a pure example of decentralized governance. Still, decentralization is a component, one exemplified by the automated penetration testing.

### *4.5. Embedded ethics*

Embedding is ethicists collaborating with engineers from the design of an AI tool and on through its deployment and continuing use.

Embedded ethics is visible in the Telus privacy practice of data minimization which limits the information requested from users to that necessary for the application's functionality [44,50]. The result is a challenge for engineers. How can the most accurate results be outputted with the least personal information inputted [51]?

Responses will vary as the subject matter shifts. A customer-support language tool responding to a technical question about internet connectivity may ask a customer about passwords and routers, while the same tool's response to a healthcare question may require data about a user's previous doctors visits. In both cases, however, derivative questions open a space shared by ethics and engineering. What constitutes minimally required information? Which parts should be categorized as sensitive? For example, the make and model of a user's router seems very different from the specialty of a doctor that has been visited, but what is the difference, exactly? And how is it measured? What does privacy actually mean in one and the other context?

Embedding ethics within engineering allows these questions to be answered as they arise and as part of the technical development. Ethics integrates into *how* research advances. So, it is not just that technology is sparking ethical questions, it is also that ethical questions are converting into technical challenges for engineers to meet. This is how ethics works when it is accelerationist. It is about finding ways to push technology forward by solving ethical problems, instead of posing pass or fail tests that determine what technology will be allowed to go forward.

Distinct and specific examples of embedding ethics at Telus could be traced through the company's Data Enablement plan, and its Impact Scanning exercises [52]. In both cases, forms of ethical expertise are developed within Telus technical design specifications and architectures [53].

More broadly, the ethics and engineering dynamic emerges in the following assertion from the Telus Director of Data Ethics: We challenge our developers to come up with solutions and to enhance the guardrails to create the outcomes we're looking for [48].

This is what acceleration ethics sounds like in the real world. The guardrail—the ethics of privacy—is a way of engineers *doing* innovation as opposed to a curbing restriction.

## 5. Conclusion

Acceleration AI ethics maximizes social responsibility *through* innovation, as opposed to sacrificing social responsibility for innovation, or sacrificing innovation for social responsibility. Five elements of the ethics have been listed, defined, and then illustrated in the context of the Telus customer service agent. The results are summarized in Table 2, below. While the purity of theoretical positions is blurred by real-world complexities, the conclusion is that the Telus case illustrates how the acceleration strategy transforms AI ethics into an innovation catalyst.

**Table 2.** Overlaps between acceleration AI ethics and the development of the Telus language tool.

| **Acceleration Elements at Telus** |
| --- |
| INNOVATION SOLVES INNOVATION PROBLEMS |
| Privacy vulnerabilities in the AI customer support tool are addressed by innovating a subsequent language tool. |
| INNOVATION IS INTRINSICALLY VALUABLE |
| Innovation's intrinsic value is audible in the imperative to "jump in and experiment" with language models. There is value in doing, independent from the effects of what gets done. |
| THE UNKNOWN IS ENCOURAGING |
| The "courage to innovate" at TELUS suggests a motivational aspect to the unforeseeable. Instead of being a reason to stop, not knowing is a reason to go forward. |
| DECENTRALIZATION |
| Automated red-teaming evolves governance from a tool's use and users, as opposed to governance as imposed and peremptory judgements. |
| EMBEDDED ETHICS |
| Ethicists and engineers collaborating transforms governance into an opportunity for still more innovation. |

## 6. Limitations of this paper and opportunities for further research

There are a number of limitations to this study, and each presents an opportunity for future research. First, the study captures the Telus experience retroactively, and primarily through public information as opposed to synchronously and with input gathered as the tool was developed. A more complete acceleration investigation would document work from a tool's ideation through deployment and use. The current paper aims to introduce and structure that larger study.

A further limitation resides in the paper's framing tension between innovation and safety. While acceleration ethics is one way to address the tension, other approaches derive from Kantian thought as well as the ethics of precaution [8]. Because this paper's tasks are limited to describing acceleration theoretically and illustrating how it works practically, those distinct approaches remain outside the scope of investigation. Still, a more extensive study would carefully distinguish the specific differences separating the various potential strategies for addressing the innovation and safety dilemma. It might also situate the thought within larger and established debates between deontological and consequentialist approaches to ethics.

Finally, there is the limitation of the subject scope. While large language models raise AI ethics challenges today, they do not represent every challenge, and it would be a mistake to blindly apply the lessons gleaned from the Telus case elsewhere. There are differences between AI language tools and retrieval tools, for example, and there are gaping divisions between the Telus customer support AI, and agents employed in hiring, and for the scoring of loan applications. Different AI tools in distinct contexts will require their own investigations of acceleration ethics.

## Declaration of generative AI and AI-assisted technologies

The authors did not use generative AI or AI-assisted technologies in the writing of this manuscript.

## Acknowledgements

Jesslyn Dymond, the Director of AI Governance and Data Ethics at TELUS, supported this paper from its conception with discussion and statements describing the development of the TELUS AI tool. Yoelit Lipinsky, PhD, Data Ethicist at TELUS, provided feedback and directions for investigation as this paper was composed. Alex Latulipe Loiselle, Technology Architect in the Data and Trust Office at TELUS, provided feedback as this paper was composed.

## Conflicts of interest

The author declares no conflicts of interest.